\documentclass[aps,prl,twocolumn,superscriptaddress,nofootinbib,amsmath,floatfix
]{revtex4-1}

\usepackage{graphicx}
\usepackage{bm}
\usepackage{natbib}
\usepackage{color}
\def\nat{Nature\ }
\def\aap{Astron.\ Astrophys.\ }
\def\apj{Astrophys.\ J.\ }
\def\apjl{Astrophys.\ J.\ Lett.\ }

\def\mnras{Mon.\ Not.\ Roy.\ Astron.\ Soc.\ }

\def\prd{Phys.\ Rev.\ D\ }
\def\prl{Phys.\ Rev.\ Lett.\ }

\def\jcap{J.\ Cosmol.\ Astropart.\ Phys.\ }

\begin{document}

\title{Klein-Nishina effect and the cosmic ray electron spectrum}

\author{Kun Fang}
\affiliation{Key Laboratory of Particle Astrophysics, Institute of High 
Energy Physics, Chinese Academy of Sciences, Beijing 100049, China}

\author{Xiao-Jun Bi}
\email{bixj@mail.ihep.ac.cn}
\affiliation{Key Laboratory of Particle Astrophysics, Institute of High 
Energy Physics, Chinese Academy of Sciences, Beijing 100049, China}
\affiliation{University of Chinese Academy of Sciences, Beijing 100049, China}

\author{Su-Jie Lin}
\affiliation{School of Physics and Astronomy, Sun Yat-Sen University, Zhuhai 
519082, China}

\author{Qiang Yuan}
\email{yuanq@pmo.ac.cn}
\affiliation{Key Laboratory of Dark Matter and Space Astronomy, Purple Mountain 
Observatory, Chinese Academy of Sciences, Nanjing 210023, China}
\affiliation{School of Astronomy and Space Science, University of Science and
Technology of China, Hefei 230026, China}
\affiliation{Center for High Energy Physics, Peking University, Beijing 100871, 
China}

\begin{abstract}
Radiative energy losses are very important in regulating the cosmic ray 
electron and/or positron (CRE) spectrum during their propagation in the Milky 
Way. Particularly, the Klein-Nishina (KN) effect of the inverse Compton 
scattering (ICS) results in less efficient energy losses of high-energy 
electrons, which is expected to leave imprints on the propagated electron 
spectrum. It has been proposed that the hardening of CRE spectra around 50 GeV 
observed by Fermi-LAT, AMS-02, and DAMPE could be due to the KN effect. We show 
in this work that the transition from the Thomson regime to the KN regime of the 
ICS is actually quite smooth compared with the approximate treatment adopted in 
some previous works. As a result, the observed spectral hardening of CREs cannot 
be explained by the KN effect. It means that an additional hardening of the 
primary electrons spectrum is needed. We also provide a parameterized form for 
the accurate calculation of the ICS energy-loss rate in a wide energy range.
\end{abstract}

\maketitle

{\it Introduction.---}Precise measurements of the energy spectra of cosmic ray 
electrons and/or positrons (CREs) have achieved big progresses in recent years
\cite{2008Natur.456..362C,2008PhRvL.101z1104A,2009PhRvL.102r1101A,
2011PhRvL.106t1101A,2014PhRvL.113v1102A,2017Natur.552...63D,
2017PhRvL.119r1101A,2017PhRvD..95h2007A,2019PhRvL.122d1102A,
2019PhRvL.122j1101A}, which are very important in probing the origin 
and propagation of CREs, as well as new physics such as the annihilation
of dark matter. Several features have been revealed in the energy spectra 
of CREs, including a softening around several GeV \cite{2011PhRvL.106t1101A,
2014PhRvL.113v1102A}, a hardening at $\sim50$~GeV, and a softening again at 
$E\sim0.9$~TeV \cite{2008PhRvL.101z1104A,2017Natur.552...63D}.
These features have interesting and important implications on the 
origin of CREs. Together with the positron excesses 
\cite{2009Natur.458..607A,2013PhRvL.110n1102A},
a three-component scenario of electrons and positrons is generally
established, which includes the {\it primary electrons} accelerated
by CR sources such as supernova remnants (SNRs), the {\it secondary electrons 
and positrons} produced by the inelastic collisions between CR nuclei and the 
interstellar medium, and the {\it additional electrons and positrons} 
accounting for the high-energy excesses (see e.g., \cite{2018SCPMA..61j1002Y}).

One prominent effect of the CRE propagation in the Milky Way is the radiative 
energy losses. The synchrotron and inverse Compton scattering (ICS) losses are 
dominant for CREs with energies $\gtrsim$ GeV 
\cite{1995PhRvD..52.3265A,1998ApJ...509..212S}, which result in an energy-loss 
rate with $\dot{E}\propto E^2$. However, the $E^2$ form for the ICS process is
only valid in the Thomson regime, when $4E\epsilon/(m_ec^2)^2 \ll 1$, where 
$\epsilon$ is the energy of the target photon, $m_e$ is the mass of electron, 
and $c$ is the speed of light. At higher energies (either the CRE or the target 
photon has a high enough energy), the ICS cross section takes the full 
Klein-Nishina (KN) form, which gets suppressed compared with the Thomson cross 
section, resulting in a smaller energy-loss rate. The KN effect is expected to 
be important even for CREs below TeV energies, from the scattering with the 
ultraviolet(UV)-optical and infrared components of the interstellar radiation 
field (ISRF). The reduction of the ICS loss rate is expected to give higher 
equilibrium CRE fluxes, leaving more complex features on the CRE spectrum than 
the case of Thomson approximation \cite{1970RvMP...42..237B}. 

Recently, Ref.~\cite{2020PhRvL.125e1101E} argued that the $\sim50$~GeV 
hardening of the CRE spectrum can be fully explained with the KN effect. 
However, the KN correction of the ICS loss adopted in 
Ref.~\cite{2020PhRvL.125e1101E} is a simple analytical approximation 
\cite{2010NJPh...12c3044S}, which is not accurate enough to correctly describe 
fine structures of the CRE spectrum. Here we will show that considering the 
exact form of the KN cross section of the ICS, the transition from the Thomson 
regime to the KN regime is much smoother than the analytical approach of 
Ref.~\cite{2010NJPh...12c3044S}. Taking the realistic ISRF distribution in the 
Milky Way into account, we will show that the observed spectral break of the 
CRE spectrum at $\sim50$~GeV by Fermi-LAT, AMS-02, and DAMPE should not be 
explained by the KN loss effect. Therefore, to properly account for the 
observed spectral feature, a hardening of the primary electron spectrum is 
still necessary
\cite{2013PhLB..727....1Y,2013PhRvD..88b3013C,2015PhRvD..91f3508L,
2015PhLB..749..267L}. 

{\it Energy-loss rate of ICS.---}Due to the efficient radiative energy losses, 
high energy CREs can only travel a short distance in the Milky Way. Typically 
for CREs above $\sim1$ GeV, the energy-loss effect dominates over the escape 
effect and becomes essential for determining the spectral shape of the 
propagated CREs. The synchrotron radiation and ICS dominate the energy losses 
of high-energy CREs. If both the accelerated electron spectrum and the energy 
dependence of the energy-loss rate have power-law forms and the CRE sources are 
homogeneously distributed at an infinite thin disk, the propagated primary 
electron spectrum should also be a power-law in high energy
\cite{2004ApJ...601..340K,2010A&A...524A..51D}. However, for the ICS, the KN 
suppression of the cross section leads to correction of the conventional $E^2$ 
form of the energy-loss rate in the Thomson limit. As a result, the propagated 
CRE spectrum should deviate from a simple power-law, as has been studied in 
many works \cite{2004ApJ...601..340K,2010ApJ...710..236S,2010NJPh...12c3044S,
2010A&A...524A..51D,2012ApJ...751...71B,2020PhRvL.125e1101E}.

Using the Lorentz factor $\gamma=E/(m_ec^2)$ as variable, the ICS 
energy-loss rate of a single electron with energy $E$ can be written as 
\cite{2010NJPh...12c3044S}
\begin{equation}
 |\dot{\gamma}|_{\rm ic}=\frac{12c\sigma_T\gamma^2}{m_ec^2}\int_0^\infty 
d\epsilon\,\epsilon\,n(\epsilon)\int_0^1dq\,\frac{qF(\Gamma, q)}{(1+\Gamma 
q)^3}\,,
 \label{eq:ic_rate}
\end{equation}
where $\sigma_T$ is the Thomson cross section, $\epsilon$ is the energy 
of the target photon, $n(\epsilon)$ is the differential number density 
of an isotropic photon field, $\Gamma=4\epsilon\gamma/(m_ec^2)$, and
\begin{equation}
F(\Gamma, q)=2q\ln q+(1+2q)(1-q)+\frac{(\Gamma q)^2(1-q)}{2(1+\Gamma q)}\,,
\label{eq:F}
\end{equation}
which is derived from the exact KN formula 
\cite{1968PhRv..167.1159J,1970RvMP...42..237B}. 

For electrons propagating in the Milky Way, the target photon field includes 
the ISRF and the cosmic microwave background (CMB). We adopt the ISRF 
averaged over 2 kpc around the Sun as a benchmark 
\cite{2010A&A...524A..51D}, which consists of five gray-body components, with 
temperatures of 23209.0 K, 6150.4 K, 3249.3 K, 313.3 K, 33.1 K, and energy 
densities of 0.12 eV cm$^{-3}$, 
0.23 eV cm$^{-3}$, 0.37 eV cm$^{-3}$, 0.055 eV cm$^{-3}$, 0.25 eV cm$^{-3}$,
respectively. The temperature and energy density of the CMB are 
2.725 K and 0.26 eV cm$^{-3}$ \cite{2009ApJ...707..916F}. 

\begin{figure}[t]
\centering
\includegraphics[width=0.48\textwidth]{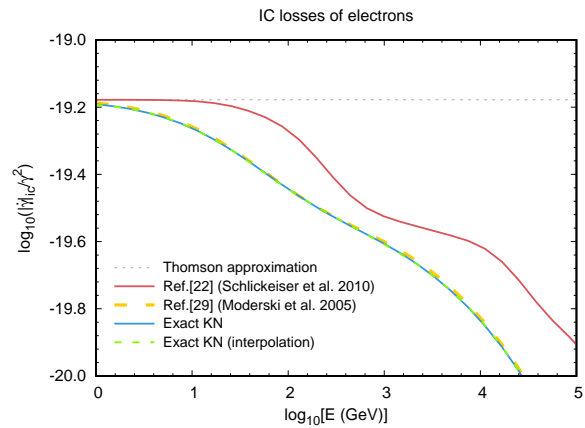}
\caption{Electron energy-loss rates due to the ICS. The blue solid line 
shows the numerical integration of Eq.~(1), and the green dashed line
is obtained from the parameterization of Eq.~(5). For comparison, we 
show the results of the Thomson regime in gray dotted line, and of the 
approximations in Refs.~\cite{2010NJPh...12c3044S,2005MNRAS.363..954M} 
in red solid line and orange dashed line, respectively. For all cases the ISRF 
is the same as that from Ref.~\cite{2010A&A...524A..51D}.}
\label{fig:ic}
\end{figure}

We show the exact ICS loss rate from a numerical integration of Eq.~(1) in Fig. 
\ref{fig:ic} with blue solid line. Compared with the result computed in the 
Thomson limit, the KN effect appears for $E\gtrsim1$~GeV. We also compare the 
ICS loss rate obtained with the approximated KN correction given in 
Ref.~\cite{2010NJPh...12c3044S}. It is shown that there is remarkable 
difference of the approximation from the numerical computation. Particularly, 
the numerical result shows less prominent features due to the KN scattering off 
different ISRF components. The transition from the Thomson regime to the 
extreme KN regime is actually broader and shallower than the approximation.

Eq.~(1) needs a two-dimensional numerical integration, and is inconvenient
for general-purpose use. We therefore try to find a proper parameterized
approximation of the exact result. For a gray-body photon field with 
temperature $T$ and energy density $w$, Eq.~(\ref{eq:ic_rate}) can be 
rewritten as
\begin{equation}
 |\dot{\gamma}|_{\rm ic}=\frac{20c\sigma_Tw\gamma^2}{\pi^4m_ec^2}Y(\gamma, T)\,,
 \label{eq:gray_body}
\end{equation}
where
\begin{equation}
 \begin{aligned}
 Y(\gamma, T)=\frac{9}{(kT)^4}&\int_0^\infty 
d\epsilon\,\frac{\epsilon^3}{{\rm 
exp}[\epsilon/(kT)]-1} \\
 &\times\int_0^1dq\,\frac{qF(\Gamma , q)}{(1+\Gamma 
q)^3}\,,
 \end{aligned}
 \label{eq:Y}
\end{equation}
with $k$ being the Boltzmann constant. Setting $x=4\gamma kT/(m_ec^2)$,
we can easily find that the integral Eq.~(\ref{eq:Y}) depends only on
the variable $x$. It means that for a gray-body photon field, there is 
a degeneracy between $\gamma$ and $T$ for the computation of the ICS
loss rate. It is evident that $x\ll 1$ corresponds to the Thomson 
regime, while $x\gg 1$ corresponds to the extreme KN regime. We find 
that for $x<1.5\times10^{-3}$ and $x>150$, $Y(x)$ can be approximated 
by the analytical formulae in the Thomson and the KN limits within an 
accuracy of 1\%. In the intermediate regime, we use a six-order 
polynomial function in the log-log space to describe $Y(x)$. 
Then we obtain the expression of $Y(x)$ in the whole range as
\begin{equation}
 Y(x)=\left\{
\begin{aligned}
 &\frac{\pi^4}{15},  & x \le 1.5\times10^{-3} \\
 &{\rm exp}\left[\sum_{i=0}^6 c_i(\ln x)^i\right], & 1.5\times10^{-3}<x<150 \\
 &\frac{3}{4}\left(\frac{\pi}{x}\right)^2(\ln x-1.9805), & x \ge 150
\end{aligned}
\right.\,.
\label{eq:Y_interp}
\end{equation}
The coefficients of the polynomial function are listed in Table 
\ref{tab:coeffi}. This approach ensures an accuracy of $<1\%$ in the 
whole energy range. The ICS loss rate calculated with 
Eq.~(\ref{eq:Y_interp}) is also shown in Fig.~\ref{fig:ic}, which is 
well consistent with the numerical calculation.
 
The approximated KN correction provided by Ref.~\cite{2005MNRAS.363..954M} also 
yields an accurate ICS loss rate (yellow dashed line in Fig.~\ref{fig:ic}). 
However, numerical integration over the background photon energy is still 
needed in that approach.
 
\begin{table}[t]
 \centering
 \caption{Coefficients of the interpolating polynomial in Eq. 
 (\ref{eq:Y_interp}).}
 \begin{tabular}{ccc}
  \hline
  \hline
  $c_0$ & $c_1$ & $c_2$  \\
  \hline
  $-3.996\times10^{-2}$~~~ & $-9.100\times10^{-1}$~~~ & $-1.197\times10^{-1}$ \\
  \hline  
  $c_3$ & $c_4$ & $c_5$ \\
  \hline
  $3.305\times10^{-3}$~~~ & $1.044\times10^{-3}$~~~ & $-7.013\times10^{-5}$ \\
  \hline
  $c_6$ & & \\
  \hline
  $-9.618\times10^{-6}$~~~ & & \\
  \hline
 \end{tabular}
 \label{tab:coeffi}
\end{table}

\begin{figure}[t]
 \centering
 \includegraphics[width=0.48\textwidth]{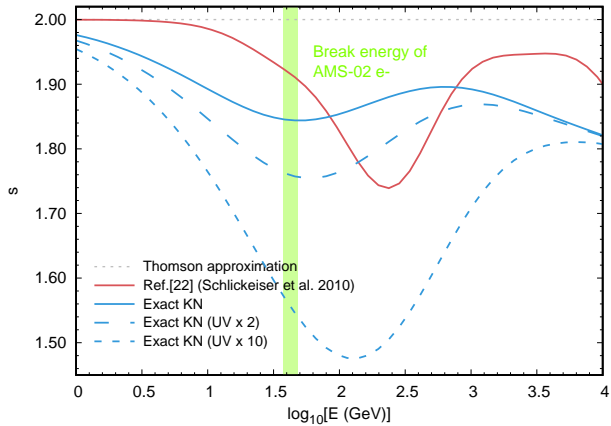}
 \caption{The slope of $\log_{10}|\dot{\gamma}|$ versus 
$\log_{10}E$ (named as $s$). This quantity reflects the impact on the 
primary electron spectrum due to the radiative losses. The blue solid line and 
the red solid line are associated with the scenarios in Fig.~\ref{fig:ic}, 
respectively. The blue dashed lines show the cases with exact KN correction 
while higher energy densities of the UV background than the default 
configuration. The green band shows the break energy (1$\sigma$ 
confidence interval) of the CRE spectrum measured by AMS-02 
\cite{2019PhRvL.122d1102A}.}
 \label{fig:b1}
\end{figure}

{\it Impact on the electron spectrum.---}Before calculating the CRE spectrum, 
we discuss the energy dependency of the energy-loss rate by 
defining
\begin{equation}
 s=\frac{d(\log_{10}|\dot{\gamma}|)}{d(\log_{10}E)}\,.
 \label{eq:slope}
\end{equation}
We would like to highlight that this quantity directly reflects the impact on 
the CRE spectrum due to the radiative losses: The decrease of $s$ corresponds 
to hardening of the CRE spectrum and vice versa. The total energy-loss rate for 
high-energy CREs is 
$|\dot{\gamma}|=|\dot{\gamma}|_{\rm ic} + |\dot{\gamma}|_{\rm syn}$, 
where $|\dot{\gamma}|_{\rm syn}$ is the synchrotron loss rate:
\begin{equation}
 |\dot{\gamma}|_{\rm syn}=\frac{\sigma_TcB^2\gamma^2}{6\pi m_ec^2}\,.
\end{equation}
We take $B=3$ $\mu$G as a benchmark interstellar magnetic field strength. The 
proportion of the synchrotron component in the energy-loss term may also 
affect the spectral features of CREs, as pointed out in 
Ref.~\cite{1985Ap.....23..650A}. However, given the typical magnetic field 
strength of the Milky Way, the synchrotron loss may only be important at much 
higher energies (for example, $\sim10$ TeV).

Fig.~\ref{fig:b1} shows the quantity $s$ calculated using the 
exact KN correction (blue solid line) and the approximated approach given in 
Refs.~\cite{2010NJPh...12c3044S} (red solid line). It can be seen that the 
approximated approach gives significantly different features. The latest 
measurement of the electron spectrum by AMS-02 indicates a spectral hardening 
at $\approx42.1$~GeV \cite{2019PhRvL.122j1101A}. It has been proposed that this 
hardening feature can be explained as the decrease of the ICS loss rate due to 
the KN effect on the UV backgrounds \cite{2020PhRvL.125e1101E}. The reason 
should be ascribed to the use of the approximation given by 
Ref.~\cite{2010NJPh...12c3044S}. Under this approximation, $s$ 
has a sharp drop at $\sim50$ GeV, which coincidentally results in a good 
accommodation to the CRE spectral feature. However, as indicated by the exact 
calculation, the KN effect actually appears at much lower energies. 
We find that the exact $Y(x)$ in Eq.~(\ref{eq:Y_interp}) deviates 
from the result of the Thomson approximation by $10\%$ when $x$ reaches 
$\sim$0.02. For typical energy of UV photons ($\sim$1 eV), this criteria 
corresponds to $E\sim0.15$ GeV. The quantity $s$ keeps 
decreasing to about 50~GeV, and then increases for $E\gtrsim50$~GeV. 
Physically, the UV component of the ICS loss rate begins to lose 
the dominance at $\sim$50~GeV, where the KN effects for the infrared and CMB 
components are still not significant. Thus, in the range from $\sim$50~GeV to 
$\sim$700~GeV, $s$ increases due to the dominance of the infrared and CMB 
components.  It means that the primary electron spectrum should become softer 
rather than harder at 50~GeV.

For the energy range we are interested in, the KN effect on the UV 
background is the main factor of the spectral variation due to the radiative 
energy losses, so we adjust the energy density of the UV component to discuss 
the impact from the uncertainties in the ISRF model. In Fig.~\ref{fig:b1}, we 
show the cases with a two times and a ten times stronger UV component 
compared with that given by Ref.~\cite{2010A&A...524A..51D}, respectively. For 
the ``UV$\times$2'' case, the KN effect still cannot predict spectral hardening 
above $\sim50$~GeV. Although there is spectral hardening above $\sim50$~GeV in 
the ``UV$\times$10'' case, the hardening effect begins at much lower energies 
due to the nature of KN correction, which obviously cannot explain the observed 
spectral break at $\sim50$~GeV.

In Ref.~\cite{2020arXiv201011955E}, a follow-up work of 
Ref.~\cite{2020PhRvL.125e1101E}, the authors argued that the CRE spectral 
hardening at $\sim$50~GeV can still be explained by the KN effect if the UV 
component of the ISRF is two times stronger than that used in 
Ref.~\cite{2020PhRvL.125e1101E}. However, after checking the left panel of 
their Fig.~11, we find that the spectrum of the ``SNR+secondary'' component 
becomes harder for $E\lesssim50$~GeV and then softer for $E\gtrsim50$~GeV, 
which is consistent with that predicted by the ``UV$\times$2'' case of the 
present work.  The hardening above $\sim$50~GeV in the total electron spectrum 
of their Fig.~11 is very likely due to the hard component from the pulsar wind 
nebulae, which does not support their own argument.

\begin{figure}[t]
 \centering
 \includegraphics[width=0.48\textwidth]{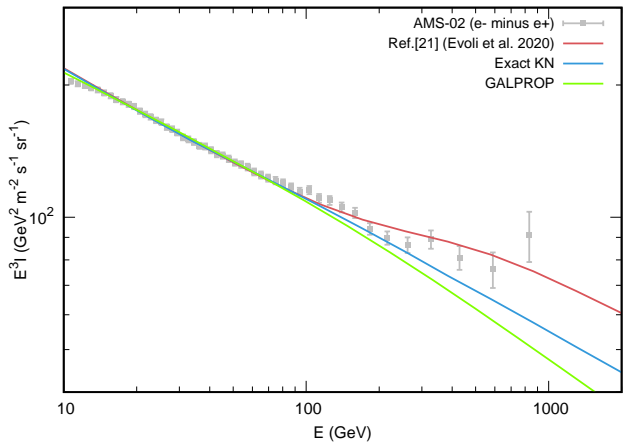}
 \caption{Propagated primary electron fluxes $I=c/(4\pi)N$ considering the 
synchrotron and ICS energy-loss effects. The red line is obtained with the ICS 
loss rate used in Ref.~\cite{2020PhRvL.125e1101E}, while the blue line is 
calculated with the exact KN correction for the ICS loss rate. 
The spectrum obatined with GALPROP is shown with the green line. We adjust each 
injection spectra index to fit the electron-minus-positron 
spectrum of AMS-02 in low energy 
\cite{2019PhRvL.122j1101A,2019PhRvL.122d1102A}.}
 \label{fig:spec}
\end{figure}

Now we discuss the CRE spectrum after propagation. First we adopt a simple
semi-analytical model to compare the effects of the different KN corrections. 
The propagation of CREs in the Galaxy can be expressed by the 
diffusion-energy-loss equation:
\begin{equation}
 \frac{\partial N}{\partial t} - D\Delta N + \frac{\partial}{\partial
E}(bN) = Q(E, \bm{x}, t) \,,
 \label{eq:prop}
\end{equation}
where $N(E)$ is the differential number density of CREs, $D(E)$ is the
diffusion coefficient, $b(E)=\dot{\gamma}m_ec^2$ is the energy-loss rate, 
and $Q(E, \bm{x}, t)$ is the source function. The diffusion coefficient 
is adopted as $D(E)=3.67\times10^{28}(E/1\,{\rm GeV})^{0.33}$, which 
is consistent with the B/C data (e.g., Ref.~\cite{2017PhRvD..95h3007Y}). 
We assume a power-law source spectrum as $Q\propto E^{-\alpha}$. 
For the spatial distribution of the CRE sources, we adopt the radial 
distribution of SNRs \cite{2015MNRAS.454.1517G} and assume all the sources are 
located at an infinite thin disk. The rate of the Galactic supernova explosion 
is assumed to be 3 per century. We obtain the stationary solution of Eq. 
(\ref{eq:prop}) using the Green's function method. A cylinder geometry of the 
diffusion zone is assumed. The half-height of the cylinder is set to be 5 kpc 
\cite{2017PhRvD..95h3007Y}. 

The propagated primary electron spectra are given in Fig.~\ref{fig:spec}. 
We also show the electron-minus-positron spectrum of AMS-02 which 
is the proper amount for the comparison with the calculated primary electron 
spectra \cite{2015PhLB..749..267L} (the flux difference between the secondary 
electrons and positrons is negligible compared with that of the primary 
electrons). The blue solid line and the red solid line are associated with the 
cases with the same color in Fig.~\ref{fig:ic} (or Fig.~\ref{fig:b1}), 
respectively. The latter is obtained with the identical ICS loss rate used in 
Ref.~\cite{2020PhRvL.125e1101E}. We adjust the injection spectral indices so 
that these two spectra can both fit the low energy data ($\alpha=2.69$ for 
the former and $\alpha=2.63$ for the latter). The difference is significant in 
high energy. It is entirely due to the different KN corrections of these two 
methods: the KN correction used by Ref.~\cite{2020PhRvL.125e1101E} predicts 
spectral hardening above $\sim50$~GeV, while the exact KN correction predicts 
spectral softening.

We also show the primary electron spectrum calculated by 
GALPROP\footnote{https://galprop.stanford.edu/} in Fig.~\ref{fig:spec}. GALPROP 
is an ideal tool for the calculation of the electron spectrum as it considers 
the spatial dependency of the ISRF, which should be the more realistic picture 
than the homogeneous ISRF used above. We adopt the same diffusion coefficient 
and height of the diffusion zone as the above calculations and assume a single 
power-law injection spectrum with $\alpha=2.73$. As can be seen, the result of 
GALPROP indicates further spectral softening in high energy, which is caused by 
the thickness of the source distribution in the vertical direction and also the 
spatially dependent ISRF. Therefore, an extra hardening for the primary 
electron spectrum is needed to fit the data as pointed out by the previous 
works using GALPROP 
\cite{2013PhLB..727....1Y,2013PhRvD..88b3013C,2015PhRvD..91f3508L,
2015PhLB..749..267L}.

{\it Conclusion.---} The radiative energy loss is crucial to 
determining the spectra of CREs during their propagation. Due to the KN effect 
of the ICS, the CRE spectrum is expected to deviate from a simple power-law 
form. We note that in some works the approximation of the ICS energy-loss rate 
in the intermediate range between the Thomson and KN limits is not accurate 
enough, which may lead to the incorrect explanation to the spectral hardening 
of CREs at $\sim50$~GeV found by several experiments. We show that the 
transition from the Thomson regime to the KN regime is actually broader and 
shallower under the exact calculation. Due to the nature of KN correction, the 
observed spectral break cannot be explained with the KN effect, even 
considering the uncertainties in the ISRF. For the convenience of practical 
use, we further give a polynomial form to describe the ICS loss rate in the 
intermediate region between the Thomson and extreme KN limits.

The source of the positron excess should contribute partly to the observed CRE 
spectral hardening, as it is assumed to generate equal amount of electrons. 
To remove the effect of the extra positron source, we compare the calculated 
primary electron spectrum with the electron-minus-positron spectrum of AMS-02. 
As the calculated spectrum keeps softening above several tens of GeV, a 
spectral hardening of the primary electron spectrum is needed to fit the data. 
Physically the origin could be due to the spectral fluctuation brought by the 
discrete distribution of cosmic ray acceleration sources (e.g.,
\cite{1970ApJ...162L.181S,1995A&A...294L..41A,2014JCAP...04..006D,
2017ApJ...836..172F}). The precise measurements of the energy spectra of 
electrons and positrons thus do reveal the properties of the CRE sources.

\acknowledgments
This work is supported by the National Key Research and Development 
Program of China (No. 2016YFA0400203, 2016YFA0400204), the National 
Natural Science Foundation of China (Nos. 11722328, U1738205, 
U1738203, 11851303, 11851305), the 100 Talents program of Chinese Academy of 
Sciences, and the Program for Innovative Talents and Entrepreneur in Jiangsu.

\bibliography{compton}

\begin{thebibliography}{37}%
\makeatletter
\providecommand \@ifxundefined [1]{%
 \@ifx{#1\undefined}
}%
\providecommand \@ifnum [1]{%
 \ifnum #1\expandafter \@firstoftwo
 \else \expandafter \@secondoftwo
 \fi
}%
\providecommand \@ifx [1]{%
 \ifx #1\expandafter \@firstoftwo
 \else \expandafter \@secondoftwo
 \fi
}%
\providecommand \natexlab [1]{#1}%
\providecommand \enquote  [1]{``#1''}%
\providecommand \bibnamefont  [1]{#1}%
\providecommand \bibfnamefont [1]{#1}%
\providecommand \citenamefont [1]{#1}%
\providecommand \href@noop [0]{\@secondoftwo}%
\providecommand \href [0]{\begingroup \@sanitize@url \@href}%
\providecommand \@href[1]{\@@startlink{#1}\@@href}%
\providecommand \@@href[1]{\endgroup#1\@@endlink}%
\providecommand \@sanitize@url [0]{\catcode `\\12\catcode `\$12\catcode
  `\&12\catcode `\#12\catcode `\^12\catcode `\_12\catcode `\%12\relax}%
\providecommand \@@startlink[1]{}%
\providecommand \@@endlink[0]{}%
\providecommand \url  [0]{\begingroup\@sanitize@url \@url }%
\providecommand \@url [1]{\endgroup\@href {#1}{\urlprefix }}%
\providecommand \urlprefix  [0]{URL }%
\providecommand \Eprint [0]{\href }%
\providecommand \doibase [0]{http://dx.doi.org/}%
\providecommand \selectlanguage [0]{\@gobble}%
\providecommand \bibinfo  [0]{\@secondoftwo}%
\providecommand \bibfield  [0]{\@secondoftwo}%
\providecommand \translation [1]{[#1]}%
\providecommand \BibitemOpen [0]{}%
\providecommand \bibitemStop [0]{}%
\providecommand \bibitemNoStop [0]{.\EOS\space}%
\providecommand \EOS [0]{\spacefactor3000\relax}%
\providecommand \BibitemShut  [1]{\csname bibitem#1\endcsname}%
\let\auto@bib@innerbib\@empty
\bibitem [{\citenamefont {{Chang}}\ \emph {et~al.}(2008)\citenamefont
  {{Chang}}, \citenamefont {{Adams}}, \citenamefont {{Ahn}} \emph
  {et~al.}}]{2008Natur.456..362C}%
  \BibitemOpen
  \bibfield  {author} {\bibinfo {author} {\bibfnamefont {J.}~\bibnamefont
  {{Chang}}}, \bibinfo {author} {\bibfnamefont {J.~H.}\ \bibnamefont
  {{Adams}}}, \bibinfo {author} {\bibfnamefont {H.~S.}\ \bibnamefont {{Ahn}}},
  \emph {et~al.},\ }\href {\doibase 10.1038/nature07477} {\bibfield  {journal}
  {\bibinfo  {journal} {\nat}\ }\textbf {\bibinfo {volume} {456}},\ \bibinfo
  {pages} {362} (\bibinfo {year} {2008})}\BibitemShut {NoStop}%
\bibitem [{\citenamefont {{Aharonian}}\ \emph {et~al.}(2008)\citenamefont
  {{Aharonian}}, \citenamefont {{Akhperjanian}}, \citenamefont {{Barres de
  Almeida}} \emph {et~al.}}]{2008PhRvL.101z1104A}%
  \BibitemOpen
  \bibfield  {author} {\bibinfo {author} {\bibfnamefont {F.}~\bibnamefont
  {{Aharonian}}}, \bibinfo {author} {\bibfnamefont {A.~G.}\ \bibnamefont
  {{Akhperjanian}}}, \bibinfo {author} {\bibfnamefont {U.}~\bibnamefont
  {{Barres de Almeida}}},  \emph {et~al.},\ }\href {\doibase
  10.1103/PhysRevLett.101.261104} {\bibfield  {journal} {\bibinfo  {journal}
  {\prl}\ }\textbf {\bibinfo {volume} {101}},\ \bibinfo {eid} {261104}
  (\bibinfo {year} {2008})},\ \Eprint {http://arxiv.org/abs/0811.3894}
  {arXiv:0811.3894 [astro-ph]} \BibitemShut {NoStop}%
\bibitem [{\citenamefont {{Abdo}}\ \emph {et~al.}(2009)\citenamefont {{Abdo}},
  \citenamefont {{Ackermann}}, \citenamefont {{Ajello}} \emph
  {et~al.}}]{2009PhRvL.102r1101A}%
  \BibitemOpen
  \bibfield  {author} {\bibinfo {author} {\bibfnamefont {A.~A.}\ \bibnamefont
  {{Abdo}}}, \bibinfo {author} {\bibfnamefont {M.}~\bibnamefont {{Ackermann}}},
  \bibinfo {author} {\bibfnamefont {M.}~\bibnamefont {{Ajello}}},  \emph
  {et~al.},\ }\href {\doibase 10.1103/PhysRevLett.102.181101} {\bibfield
  {journal} {\bibinfo  {journal} {\prl}\ }\textbf {\bibinfo {volume} {102}},\
  \bibinfo {eid} {181101} (\bibinfo {year} {2009})},\ \Eprint
  {http://arxiv.org/abs/0905.0025} {arXiv:0905.0025 [astro-ph.HE]} \BibitemShut
  {NoStop}%
\bibitem [{\citenamefont {{Adriani}}\ \emph {et~al.}(2011)\citenamefont
  {{Adriani}}, \citenamefont {{Barbarino}}, \citenamefont {{Bazilevskaya}}
  \emph {et~al.}}]{2011PhRvL.106t1101A}%
  \BibitemOpen
  \bibfield  {author} {\bibinfo {author} {\bibfnamefont {O.}~\bibnamefont
  {{Adriani}}}, \bibinfo {author} {\bibfnamefont {G.~C.}\ \bibnamefont
  {{Barbarino}}}, \bibinfo {author} {\bibfnamefont {G.~A.}\ \bibnamefont
  {{Bazilevskaya}}},  \emph {et~al.},\ }\href {\doibase
  10.1103/PhysRevLett.106.201101} {\bibfield  {journal} {\bibinfo  {journal}
  {\prl}\ }\textbf {\bibinfo {volume} {106}},\ \bibinfo {eid} {201101}
  (\bibinfo {year} {2011})},\ \Eprint {http://arxiv.org/abs/1103.2880}
  {arXiv:1103.2880 [astro-ph.HE]} \BibitemShut {NoStop}%
\bibitem [{\citenamefont {{Aguilar}}\ \emph {et~al.}(2014)\citenamefont
  {{Aguilar}}, \citenamefont {{Aisa}}, \citenamefont {{Alpat}} \emph
  {et~al.}}]{2014PhRvL.113v1102A}%
  \BibitemOpen
  \bibfield  {author} {\bibinfo {author} {\bibfnamefont {M.}~\bibnamefont
  {{Aguilar}}}, \bibinfo {author} {\bibfnamefont {D.}~\bibnamefont {{Aisa}}},
  \bibinfo {author} {\bibfnamefont {B.}~\bibnamefont {{Alpat}}},  \emph
  {et~al.},\ }\href {\doibase 10.1103/PhysRevLett.113.221102} {\bibfield
  {journal} {\bibinfo  {journal} {\prl}\ }\textbf {\bibinfo {volume} {113}},\
  \bibinfo {eid} {221102} (\bibinfo {year} {2014})}\BibitemShut {NoStop}%
\bibitem [{\citenamefont {{Ambrosi}}\ \emph {et~al.}(2017)\citenamefont
  {{Ambrosi}}, \citenamefont {{An}}, \citenamefont {{Asfandiyarov}} \emph
  {et~al.}}]{2017Natur.552...63D}%
  \BibitemOpen
  \bibfield  {author} {\bibinfo {author} {\bibfnamefont {G.}~\bibnamefont
  {{Ambrosi}}}, \bibinfo {author} {\bibfnamefont {Q.}~\bibnamefont {{An}}},
  \bibinfo {author} {\bibfnamefont {R.}~\bibnamefont {{Asfandiyarov}}},  \emph
  {et~al.},\ }\href {\doibase 10.1038/nature24475} {\bibfield  {journal}
  {\bibinfo  {journal} {\nat}\ }\textbf {\bibinfo {volume} {552}},\ \bibinfo
  {pages} {63} (\bibinfo {year} {2017})},\ \Eprint
  {http://arxiv.org/abs/1711.10981} {arXiv:1711.10981 [astro-ph.HE]}
  \BibitemShut {NoStop}%
\bibitem [{\citenamefont {{Adriani}}\ \emph {et~al.}(2017)\citenamefont
  {{Adriani}}, \citenamefont {{Akaike}}, \citenamefont {{Asano}} \emph
  {et~al.}}]{2017PhRvL.119r1101A}%
  \BibitemOpen
  \bibfield  {author} {\bibinfo {author} {\bibfnamefont {O.}~\bibnamefont
  {{Adriani}}}, \bibinfo {author} {\bibfnamefont {Y.}~\bibnamefont {{Akaike}}},
  \bibinfo {author} {\bibfnamefont {K.}~\bibnamefont {{Asano}}},  \emph
  {et~al.},\ }\href {\doibase 10.1103/PhysRevLett.119.181101} {\bibfield
  {journal} {\bibinfo  {journal} {\prl}\ }\textbf {\bibinfo {volume} {119}},\
  \bibinfo {eid} {181101} (\bibinfo {year} {2017})},\ \Eprint
  {http://arxiv.org/abs/1712.01711} {arXiv:1712.01711 [astro-ph.HE]}
  \BibitemShut {NoStop}%
\bibitem [{\citenamefont {{Abdollahi}}\ \emph {et~al.}(2017)\citenamefont
  {{Abdollahi}}, \citenamefont {{Ackermann}}, \citenamefont {{Ajello}} \emph
  {et~al.}}]{2017PhRvD..95h2007A}%
  \BibitemOpen
  \bibfield  {author} {\bibinfo {author} {\bibfnamefont {S.}~\bibnamefont
  {{Abdollahi}}}, \bibinfo {author} {\bibfnamefont {M.}~\bibnamefont
  {{Ackermann}}}, \bibinfo {author} {\bibfnamefont {M.}~\bibnamefont
  {{Ajello}}},  \emph {et~al.},\ }\href {\doibase 10.1103/PhysRevD.95.082007}
  {\bibfield  {journal} {\bibinfo  {journal} {\prd}\ }\textbf {\bibinfo
  {volume} {95}},\ \bibinfo {eid} {082007} (\bibinfo {year} {2017})},\ \Eprint
  {http://arxiv.org/abs/1704.07195} {arXiv:1704.07195 [astro-ph.HE]}
  \BibitemShut {NoStop}%
\bibitem [{\citenamefont {{Aguilar}}\ \emph
  {et~al.}(2019{\natexlab{a}})\citenamefont {{Aguilar}}, \citenamefont {{Ali
  Cavasonza}}, \citenamefont {{Ambrosi}} \emph {et~al.}}]{2019PhRvL.122d1102A}%
  \BibitemOpen
  \bibfield  {author} {\bibinfo {author} {\bibfnamefont {M.}~\bibnamefont
  {{Aguilar}}}, \bibinfo {author} {\bibfnamefont {L.}~\bibnamefont {{Ali
  Cavasonza}}}, \bibinfo {author} {\bibfnamefont {G.}~\bibnamefont
  {{Ambrosi}}},  \emph {et~al.},\ }\href {\doibase
  10.1103/PhysRevLett.122.041102} {\bibfield  {journal} {\bibinfo  {journal}
  {\prl}\ }\textbf {\bibinfo {volume} {122}},\ \bibinfo {eid} {041102}
  (\bibinfo {year} {2019}{\natexlab{a}})}\BibitemShut {NoStop}%
\bibitem [{\citenamefont {{Aguilar}}\ \emph
  {et~al.}(2019{\natexlab{b}})\citenamefont {{Aguilar}}, \citenamefont {{Ali
  Cavasonza}}, \citenamefont {{Alpat}} \emph {et~al.}}]{2019PhRvL.122j1101A}%
  \BibitemOpen
  \bibfield  {author} {\bibinfo {author} {\bibfnamefont {M.}~\bibnamefont
  {{Aguilar}}}, \bibinfo {author} {\bibfnamefont {L.}~\bibnamefont {{Ali
  Cavasonza}}}, \bibinfo {author} {\bibfnamefont {B.}~\bibnamefont {{Alpat}}},
  \emph {et~al.},\ }\href {\doibase 10.1103/PhysRevLett.122.101101} {\bibfield
  {journal} {\bibinfo  {journal} {\prl}\ }\textbf {\bibinfo {volume} {122}},\
  \bibinfo {eid} {101101} (\bibinfo {year} {2019}{\natexlab{b}})}\BibitemShut
  {NoStop}%
\bibitem [{\citenamefont {{Adriani}}\ \emph {et~al.}(2009)\citenamefont
  {{Adriani}}, \citenamefont {{Barbarino}}, \citenamefont {{Bazilevskaya}}
  \emph {et~al.}}]{2009Natur.458..607A}%
  \BibitemOpen
  \bibfield  {author} {\bibinfo {author} {\bibfnamefont {O.}~\bibnamefont
  {{Adriani}}}, \bibinfo {author} {\bibfnamefont {G.~C.}\ \bibnamefont
  {{Barbarino}}}, \bibinfo {author} {\bibfnamefont {G.~A.}\ \bibnamefont
  {{Bazilevskaya}}},  \emph {et~al.},\ }\href {\doibase 10.1038/nature07942}
  {\bibfield  {journal} {\bibinfo  {journal} {\nat}\ }\textbf {\bibinfo
  {volume} {458}},\ \bibinfo {pages} {607} (\bibinfo {year} {2009})},\ \Eprint
  {http://arxiv.org/abs/0810.4995} {arXiv:0810.4995 [astro-ph]} \BibitemShut
  {NoStop}%
\bibitem [{\citenamefont {{Aguilar}}\ \emph {et~al.}(2013)\citenamefont
  {{Aguilar}}, \citenamefont {{Alberti}}, \citenamefont {{Alpat}} \emph
  {et~al.}}]{2013PhRvL.110n1102A}%
  \BibitemOpen
  \bibfield  {author} {\bibinfo {author} {\bibfnamefont {M.}~\bibnamefont
  {{Aguilar}}}, \bibinfo {author} {\bibfnamefont {G.}~\bibnamefont
  {{Alberti}}}, \bibinfo {author} {\bibfnamefont {B.}~\bibnamefont {{Alpat}}},
  \emph {et~al.},\ }\href {\doibase 10.1103/PhysRevLett.110.141102} {\bibfield
  {journal} {\bibinfo  {journal} {\prl}\ }\textbf {\bibinfo {volume} {110}},\
  \bibinfo {eid} {141102} (\bibinfo {year} {2013})}\BibitemShut {NoStop}%
\bibitem [{\citenamefont {{Yuan}}\ and\ \citenamefont
  {{Feng}}(2018)}]{2018SCPMA..61j1002Y}%
  \BibitemOpen
  \bibfield  {author} {\bibinfo {author} {\bibfnamefont {Q.}~\bibnamefont
  {{Yuan}}}\ and\ \bibinfo {author} {\bibfnamefont {L.}~\bibnamefont
  {{Feng}}},\ }\href {\doibase 10.1007/s11433-018-9226-y} {\bibfield  {journal}
  {\bibinfo  {journal} {Science China Physics, Mechanics, and Astronomy}\
  }\textbf {\bibinfo {volume} {61}},\ \bibinfo {eid} {101002} (\bibinfo {year}
  {2018})},\ \Eprint {http://arxiv.org/abs/1807.11638} {arXiv:1807.11638
  [astro-ph.HE]} \BibitemShut {NoStop}%
\bibitem [{\citenamefont {{Atoyan}}\ \emph {et~al.}(1995)\citenamefont
  {{Atoyan}}, \citenamefont {{Aharonian}},\ and\ \citenamefont
  {{V{\"o}lk}}}]{1995PhRvD..52.3265A}%
  \BibitemOpen
  \bibfield  {author} {\bibinfo {author} {\bibfnamefont {A.~M.}\ \bibnamefont
  {{Atoyan}}}, \bibinfo {author} {\bibfnamefont {F.~A.}\ \bibnamefont
  {{Aharonian}}}, \ and\ \bibinfo {author} {\bibfnamefont {H.~J.}\ \bibnamefont
  {{V{\"o}lk}}},\ }\href {\doibase 10.1103/PhysRevD.52.3265} {\bibfield
  {journal} {\bibinfo  {journal} {\prd}\ }\textbf {\bibinfo {volume} {52}},\
  \bibinfo {pages} {3265} (\bibinfo {year} {1995})}\BibitemShut {NoStop}%
\bibitem [{\citenamefont {{Strong}}\ and\ \citenamefont
  {{Moskalenko}}(1998)}]{1998ApJ...509..212S}%
  \BibitemOpen
  \bibfield  {author} {\bibinfo {author} {\bibfnamefont {A.~W.}\ \bibnamefont
  {{Strong}}}\ and\ \bibinfo {author} {\bibfnamefont {I.~V.}\ \bibnamefont
  {{Moskalenko}}},\ }\href {\doibase 10.1086/306470} {\bibfield  {journal}
  {\bibinfo  {journal} {\apj}\ }\textbf {\bibinfo {volume} {509}},\ \bibinfo
  {pages} {212} (\bibinfo {year} {1998})},\ \Eprint
  {http://arxiv.org/abs/astro-ph/9807150} {arXiv:astro-ph/9807150 [astro-ph]}
  \BibitemShut {NoStop}%
\bibitem [{\citenamefont {{Blumenthal}}\ and\ \citenamefont
  {{Gould}}(1970)}]{1970RvMP...42..237B}%
  \BibitemOpen
  \bibfield  {author} {\bibinfo {author} {\bibfnamefont {G.~R.}\ \bibnamefont
  {{Blumenthal}}}\ and\ \bibinfo {author} {\bibfnamefont {R.~J.}\ \bibnamefont
  {{Gould}}},\ }\href {\doibase 10.1103/RevModPhys.42.237} {\bibfield
  {journal} {\bibinfo  {journal} {Reviews of Modern Physics}\ }\textbf
  {\bibinfo {volume} {42}},\ \bibinfo {pages} {237} (\bibinfo {year}
  {1970})}\BibitemShut {NoStop}%
\bibitem [{\citenamefont {{Evoli}}\ \emph
  {et~al.}(2020{\natexlab{a}})\citenamefont {{Evoli}}, \citenamefont {{Blasi}},
  \citenamefont {{Amato}},\ and\ \citenamefont
  {{Aloisio}}}]{2020PhRvL.125e1101E}%
  \BibitemOpen
  \bibfield  {author} {\bibinfo {author} {\bibfnamefont {C.}~\bibnamefont
  {{Evoli}}}, \bibinfo {author} {\bibfnamefont {P.}~\bibnamefont {{Blasi}}},
  \bibinfo {author} {\bibfnamefont {E.}~\bibnamefont {{Amato}}}, \ and\
  \bibinfo {author} {\bibfnamefont {R.}~\bibnamefont {{Aloisio}}},\ }\href
  {\doibase 10.1103/PhysRevLett.125.051101} {\bibfield  {journal} {\bibinfo
  {journal} {\prl}\ }\textbf {\bibinfo {volume} {125}},\ \bibinfo {eid}
  {051101} (\bibinfo {year} {2020}{\natexlab{a}})},\ \Eprint
  {http://arxiv.org/abs/2007.01302} {arXiv:2007.01302 [astro-ph.HE]}
  \BibitemShut {NoStop}%
\bibitem [{\citenamefont {{Schlickeiser}}\ and\ \citenamefont
  {{Ruppel}}(2010)}]{2010NJPh...12c3044S}%
  \BibitemOpen
  \bibfield  {author} {\bibinfo {author} {\bibfnamefont {R.}~\bibnamefont
  {{Schlickeiser}}}\ and\ \bibinfo {author} {\bibfnamefont {J.}~\bibnamefont
  {{Ruppel}}},\ }\href {\doibase 10.1088/1367-2630/12/3/033044} {\bibfield
  {journal} {\bibinfo  {journal} {New Journal of Physics}\ }\textbf {\bibinfo
  {volume} {12}},\ \bibinfo {eid} {033044} (\bibinfo {year} {2010})},\ \Eprint
  {http://arxiv.org/abs/0908.2183} {arXiv:0908.2183 [astro-ph.HE]} \BibitemShut
  {NoStop}%
\bibitem [{\citenamefont {{Yuan}}\ and\ \citenamefont
  {{Bi}}(2013)}]{2013PhLB..727....1Y}%
  \BibitemOpen
  \bibfield  {author} {\bibinfo {author} {\bibfnamefont {Q.}~\bibnamefont
  {{Yuan}}}\ and\ \bibinfo {author} {\bibfnamefont {X.-J.}\ \bibnamefont
  {{Bi}}},\ }\href {\doibase 10.1016/j.physletb.2013.10.010} {\bibfield
  {journal} {\bibinfo  {journal} {Physics Letters B}\ }\textbf {\bibinfo
  {volume} {727}},\ \bibinfo {pages} {1} (\bibinfo {year} {2013})},\ \Eprint
  {http://arxiv.org/abs/1304.2687} {arXiv:1304.2687 [astro-ph.HE]} \BibitemShut
  {NoStop}%
\bibitem [{\citenamefont {{Cholis}}\ and\ \citenamefont
  {{Hooper}}(2013)}]{2013PhRvD..88b3013C}%
  \BibitemOpen
  \bibfield  {author} {\bibinfo {author} {\bibfnamefont {I.}~\bibnamefont
  {{Cholis}}}\ and\ \bibinfo {author} {\bibfnamefont {D.}~\bibnamefont
  {{Hooper}}},\ }\href {\doibase 10.1103/PhysRevD.88.023013} {\bibfield
  {journal} {\bibinfo  {journal} {\prd}\ }\textbf {\bibinfo {volume} {88}},\
  \bibinfo {eid} {023013} (\bibinfo {year} {2013})},\ \Eprint
  {http://arxiv.org/abs/1304.1840} {arXiv:1304.1840 [astro-ph.HE]} \BibitemShut
  {NoStop}%
\bibitem [{\citenamefont {{Lin}}\ \emph {et~al.}(2015)\citenamefont {{Lin}},
  \citenamefont {{Yuan}},\ and\ \citenamefont {{Bi}}}]{2015PhRvD..91f3508L}%
  \BibitemOpen
  \bibfield  {author} {\bibinfo {author} {\bibfnamefont {S.-J.}\ \bibnamefont
  {{Lin}}}, \bibinfo {author} {\bibfnamefont {Q.}~\bibnamefont {{Yuan}}}, \
  and\ \bibinfo {author} {\bibfnamefont {X.-J.}\ \bibnamefont {{Bi}}},\ }\href
  {\doibase 10.1103/PhysRevD.91.063508} {\bibfield  {journal} {\bibinfo
  {journal} {\prd}\ }\textbf {\bibinfo {volume} {91}},\ \bibinfo {eid} {063508}
  (\bibinfo {year} {2015})},\ \Eprint {http://arxiv.org/abs/1409.6248}
  {arXiv:1409.6248 [astro-ph.HE]} \BibitemShut {NoStop}%
\bibitem [{\citenamefont {{Li}}\ \emph {et~al.}(2015)\citenamefont {{Li}},
  \citenamefont {{Shen}}, \citenamefont {{Lu}} \emph
  {et~al.}}]{2015PhLB..749..267L}%
  \BibitemOpen
  \bibfield  {author} {\bibinfo {author} {\bibfnamefont {X.}~\bibnamefont
  {{Li}}}, \bibinfo {author} {\bibfnamefont {Z.-Q.}\ \bibnamefont {{Shen}}},
  \bibinfo {author} {\bibfnamefont {B.-Q.}\ \bibnamefont {{Lu}}},  \emph
  {et~al.},\ }\href {\doibase 10.1016/j.physletb.2015.08.001} {\bibfield
  {journal} {\bibinfo  {journal} {Physics Letters B}\ }\textbf {\bibinfo
  {volume} {749}},\ \bibinfo {pages} {267} (\bibinfo {year} {2015})},\ \Eprint
  {http://arxiv.org/abs/1412.1550} {arXiv:1412.1550 [astro-ph.HE]} \BibitemShut
  {NoStop}%
\bibitem [{\citenamefont {{Kobayashi}}\ \emph {et~al.}(2004)\citenamefont
  {{Kobayashi}}, \citenamefont {{Komori}}, \citenamefont {{Yoshida}},\ and\
  \citenamefont {{Nishimura}}}]{2004ApJ...601..340K}%
  \BibitemOpen
  \bibfield  {author} {\bibinfo {author} {\bibfnamefont {T.}~\bibnamefont
  {{Kobayashi}}}, \bibinfo {author} {\bibfnamefont {Y.}~\bibnamefont
  {{Komori}}}, \bibinfo {author} {\bibfnamefont {K.}~\bibnamefont {{Yoshida}}},
  \ and\ \bibinfo {author} {\bibfnamefont {J.}~\bibnamefont {{Nishimura}}},\
  }\href {\doibase 10.1086/380431} {\bibfield  {journal} {\bibinfo  {journal}
  {\apj}\ }\textbf {\bibinfo {volume} {601}},\ \bibinfo {pages} {340} (\bibinfo
  {year} {2004})},\ \Eprint {http://arxiv.org/abs/astro-ph/0308470}
  {arXiv:astro-ph/0308470 [astro-ph]} \BibitemShut {NoStop}%
\bibitem [{\citenamefont {{Delahaye}}\ \emph {et~al.}(2010)\citenamefont
  {{Delahaye}}, \citenamefont {{Lavalle}}, \citenamefont {{Lineros}},
  \citenamefont {{Donato}},\ and\ \citenamefont
  {{Fornengo}}}]{2010A&A...524A..51D}%
  \BibitemOpen
  \bibfield  {author} {\bibinfo {author} {\bibfnamefont {T.}~\bibnamefont
  {{Delahaye}}}, \bibinfo {author} {\bibfnamefont {J.}~\bibnamefont
  {{Lavalle}}}, \bibinfo {author} {\bibfnamefont {R.}~\bibnamefont
  {{Lineros}}}, \bibinfo {author} {\bibfnamefont {F.}~\bibnamefont {{Donato}}},
  \ and\ \bibinfo {author} {\bibfnamefont {N.}~\bibnamefont {{Fornengo}}},\
  }\href {\doibase 10.1051/0004-6361/201014225} {\bibfield  {journal} {\bibinfo
   {journal} {\aap}\ }\textbf {\bibinfo {volume} {524}},\ \bibinfo {eid} {A51}
  (\bibinfo {year} {2010})},\ \Eprint {http://arxiv.org/abs/1002.1910}
  {arXiv:1002.1910 [astro-ph.HE]} \BibitemShut {NoStop}%
\bibitem [{\citenamefont {{Stawarz}}\ \emph {et~al.}(2010)\citenamefont
  {{Stawarz}}, \citenamefont {{Petrosian}},\ and\ \citenamefont
  {{Blandford}}}]{2010ApJ...710..236S}%
  \BibitemOpen
  \bibfield  {author} {\bibinfo {author} {\bibfnamefont {{\L}.}~\bibnamefont
  {{Stawarz}}}, \bibinfo {author} {\bibfnamefont {V.}~\bibnamefont
  {{Petrosian}}}, \ and\ \bibinfo {author} {\bibfnamefont {R.~D.}\ \bibnamefont
  {{Blandford}}},\ }\href {\doibase 10.1088/0004-637X/710/1/236} {\bibfield
  {journal} {\bibinfo  {journal} {\apj}\ }\textbf {\bibinfo {volume} {710}},\
  \bibinfo {pages} {236} (\bibinfo {year} {2010})},\ \Eprint
  {http://arxiv.org/abs/0908.1094} {arXiv:0908.1094 [astro-ph.GA]} \BibitemShut
  {NoStop}%
\bibitem [{\citenamefont {{Blies}}\ and\ \citenamefont
  {{Schlickeiser}}(2012)}]{2012ApJ...751...71B}%
  \BibitemOpen
  \bibfield  {author} {\bibinfo {author} {\bibfnamefont {P.}~\bibnamefont
  {{Blies}}}\ and\ \bibinfo {author} {\bibfnamefont {R.}~\bibnamefont
  {{Schlickeiser}}},\ }\href {\doibase 10.1088/0004-637X/751/1/71} {\bibfield
  {journal} {\bibinfo  {journal} {\apj}\ }\textbf {\bibinfo {volume} {751}},\
  \bibinfo {eid} {71} (\bibinfo {year} {2012})}\BibitemShut {NoStop}%
\bibitem [{\citenamefont {{Jones}}(1968)}]{1968PhRv..167.1159J}%
  \BibitemOpen
  \bibfield  {author} {\bibinfo {author} {\bibfnamefont {F.~C.}\ \bibnamefont
  {{Jones}}},\ }\href {\doibase 10.1103/PhysRev.167.1159} {\bibfield  {journal}
  {\bibinfo  {journal} {Physical Review}\ }\textbf {\bibinfo {volume} {167}},\
  \bibinfo {pages} {1159} (\bibinfo {year} {1968})}\BibitemShut {NoStop}%
\bibitem [{\citenamefont {{Fixsen}}(2009)}]{2009ApJ...707..916F}%
  \BibitemOpen
  \bibfield  {author} {\bibinfo {author} {\bibfnamefont {D.~J.}\ \bibnamefont
  {{Fixsen}}},\ }\href {\doibase 10.1088/0004-637X/707/2/916} {\bibfield
  {journal} {\bibinfo  {journal} {\apj}\ }\textbf {\bibinfo {volume} {707}},\
  \bibinfo {pages} {916} (\bibinfo {year} {2009})},\ \Eprint
  {http://arxiv.org/abs/0911.1955} {arXiv:0911.1955 [astro-ph.CO]} \BibitemShut
  {NoStop}%
\bibitem [{\citenamefont {{Moderski}}\ \emph {et~al.}(2005)\citenamefont
  {{Moderski}}, \citenamefont {{Sikora}}, \citenamefont {{Coppi}},\ and\
  \citenamefont {{Aharonian}}}]{2005MNRAS.363..954M}%
  \BibitemOpen
  \bibfield  {author} {\bibinfo {author} {\bibfnamefont {R.}~\bibnamefont
  {{Moderski}}}, \bibinfo {author} {\bibfnamefont {M.}~\bibnamefont
  {{Sikora}}}, \bibinfo {author} {\bibfnamefont {P.~S.}\ \bibnamefont
  {{Coppi}}}, \ and\ \bibinfo {author} {\bibfnamefont {F.}~\bibnamefont
  {{Aharonian}}},\ }\href {\doibase 10.1111/j.1365-2966.2005.09494.x}
  {\bibfield  {journal} {\bibinfo  {journal} {\mnras}\ }\textbf {\bibinfo
  {volume} {363}},\ \bibinfo {pages} {954} (\bibinfo {year} {2005})},\ \Eprint
  {http://arxiv.org/abs/astro-ph/0504388} {arXiv:astro-ph/0504388 [astro-ph]}
  \BibitemShut {NoStop}%
\bibitem [{\citenamefont {{Agaronyan}}\ and\ \citenamefont
  {{Ambartsumyan}}(1985)}]{1985Ap.....23..650A}%
  \BibitemOpen
  \bibfield  {author} {\bibinfo {author} {\bibfnamefont {F.~A.}\ \bibnamefont
  {{Agaronyan}}}\ and\ \bibinfo {author} {\bibfnamefont {A.~S.}\ \bibnamefont
  {{Ambartsumyan}}},\ }\href {\doibase 10.1007/BF01008222} {\bibfield
  {journal} {\bibinfo  {journal} {Astrophysics}\ }\textbf {\bibinfo {volume}
  {23}},\ \bibinfo {pages} {650} (\bibinfo {year} {1985})}\BibitemShut
  {NoStop}%
\bibitem [{\citenamefont {{Evoli}}\ \emph
  {et~al.}(2020{\natexlab{b}})\citenamefont {{Evoli}}, \citenamefont {{Amato}},
  \citenamefont {{Blasi}},\ and\ \citenamefont
  {{Aloisio}}}]{2020arXiv201011955E}%
  \BibitemOpen
  \bibfield  {author} {\bibinfo {author} {\bibfnamefont {C.}~\bibnamefont
  {{Evoli}}}, \bibinfo {author} {\bibfnamefont {E.}~\bibnamefont {{Amato}}},
  \bibinfo {author} {\bibfnamefont {P.}~\bibnamefont {{Blasi}}}, \ and\
  \bibinfo {author} {\bibfnamefont {R.}~\bibnamefont {{Aloisio}}},\ }\href@noop
  {} {\bibfield  {journal} {\bibinfo  {journal} {arXiv e-prints}\ ,\ \bibinfo
  {eid} {arXiv:2010.11955}} (\bibinfo {year} {2020}{\natexlab{b}})},\ \Eprint
  {http://arxiv.org/abs/2010.11955} {arXiv:2010.11955 [astro-ph.HE]}
  \BibitemShut {NoStop}%
\bibitem [{\citenamefont {{Yuan}}\ \emph {et~al.}(2017)\citenamefont {{Yuan}},
  \citenamefont {{Lin}}, \citenamefont {{Fang}},\ and\ \citenamefont
  {{Bi}}}]{2017PhRvD..95h3007Y}%
  \BibitemOpen
  \bibfield  {author} {\bibinfo {author} {\bibfnamefont {Q.}~\bibnamefont
  {{Yuan}}}, \bibinfo {author} {\bibfnamefont {S.-J.}\ \bibnamefont {{Lin}}},
  \bibinfo {author} {\bibfnamefont {K.}~\bibnamefont {{Fang}}}, \ and\ \bibinfo
  {author} {\bibfnamefont {X.-J.}\ \bibnamefont {{Bi}}},\ }\href {\doibase
  10.1103/PhysRevD.95.083007} {\bibfield  {journal} {\bibinfo  {journal}
  {\prd}\ }\textbf {\bibinfo {volume} {95}},\ \bibinfo {eid} {083007} (\bibinfo
  {year} {2017})},\ \Eprint {http://arxiv.org/abs/1701.06149} {arXiv:1701.06149
  [astro-ph.HE]} \BibitemShut {NoStop}%
\bibitem [{\citenamefont {{Green}}(2015)}]{2015MNRAS.454.1517G}%
  \BibitemOpen
  \bibfield  {author} {\bibinfo {author} {\bibfnamefont {D.~A.}\ \bibnamefont
  {{Green}}},\ }\href {\doibase 10.1093/mnras/stv1885} {\bibfield  {journal}
  {\bibinfo  {journal} {\mnras}\ }\textbf {\bibinfo {volume} {454}},\ \bibinfo
  {pages} {1517} (\bibinfo {year} {2015})},\ \Eprint
  {http://arxiv.org/abs/1508.02931} {arXiv:1508.02931 [astro-ph.HE]}
  \BibitemShut {NoStop}%
\bibitem [{\citenamefont {{Shen}}(1970)}]{1970ApJ...162L.181S}%
  \BibitemOpen
  \bibfield  {author} {\bibinfo {author} {\bibfnamefont {C.~S.}\ \bibnamefont
  {{Shen}}},\ }\href {\doibase 10.1086/180650} {\bibfield  {journal} {\bibinfo
  {journal} {\apjl}\ }\textbf {\bibinfo {volume} {162}},\ \bibinfo {pages}
  {L181} (\bibinfo {year} {1970})}\BibitemShut {NoStop}%
\bibitem [{\citenamefont {{Aharonian}}\ \emph {et~al.}(1995)\citenamefont
  {{Aharonian}}, \citenamefont {{Atoyan}},\ and\ \citenamefont
  {{Voelk}}}]{1995A&A...294L..41A}%
  \BibitemOpen
  \bibfield  {author} {\bibinfo {author} {\bibfnamefont {F.~A.}\ \bibnamefont
  {{Aharonian}}}, \bibinfo {author} {\bibfnamefont {A.~M.}\ \bibnamefont
  {{Atoyan}}}, \ and\ \bibinfo {author} {\bibfnamefont {H.~J.}\ \bibnamefont
  {{Voelk}}},\ }\href@noop {} {\bibfield  {journal} {\bibinfo  {journal}
  {\aap}\ }\textbf {\bibinfo {volume} {294}},\ \bibinfo {pages} {L41} (\bibinfo
  {year} {1995})}\BibitemShut {NoStop}%
\bibitem [{\citenamefont {{Di Mauro}}\ \emph {et~al.}(2014)\citenamefont {{Di
  Mauro}}, \citenamefont {{Donato}}, \citenamefont {{Fornengo}}, \citenamefont
  {{Lineros}},\ and\ \citenamefont {{Vittino}}}]{2014JCAP...04..006D}%
  \BibitemOpen
  \bibfield  {author} {\bibinfo {author} {\bibfnamefont {M.}~\bibnamefont {{Di
  Mauro}}}, \bibinfo {author} {\bibfnamefont {F.}~\bibnamefont {{Donato}}},
  \bibinfo {author} {\bibfnamefont {N.}~\bibnamefont {{Fornengo}}}, \bibinfo
  {author} {\bibfnamefont {R.}~\bibnamefont {{Lineros}}}, \ and\ \bibinfo
  {author} {\bibfnamefont {A.}~\bibnamefont {{Vittino}}},\ }\href {\doibase
  10.1088/1475-7516/2014/04/006} {\bibfield  {journal} {\bibinfo  {journal}
  {\jcap}\ }\textbf {\bibinfo {volume} {2014}},\ \bibinfo {eid} {006} (\bibinfo
  {year} {2014})},\ \Eprint {http://arxiv.org/abs/1402.0321} {arXiv:1402.0321
  [astro-ph.HE]} \BibitemShut {NoStop}%
\bibitem [{\citenamefont {{Fang}}\ \emph {et~al.}(2017)\citenamefont {{Fang}},
  \citenamefont {{Wang}}, \citenamefont {{Bi}}, \citenamefont {{Lin}},\ and\
  \citenamefont {{Yin}}}]{2017ApJ...836..172F}%
  \BibitemOpen
  \bibfield  {author} {\bibinfo {author} {\bibfnamefont {K.}~\bibnamefont
  {{Fang}}}, \bibinfo {author} {\bibfnamefont {B.-B.}\ \bibnamefont {{Wang}}},
  \bibinfo {author} {\bibfnamefont {X.-J.}\ \bibnamefont {{Bi}}}, \bibinfo
  {author} {\bibfnamefont {S.-J.}\ \bibnamefont {{Lin}}}, \ and\ \bibinfo
  {author} {\bibfnamefont {P.-F.}\ \bibnamefont {{Yin}}},\ }\href {\doibase
  10.3847/1538-4357/aa5b93} {\bibfield  {journal} {\bibinfo  {journal} {\apj}\
  }\textbf {\bibinfo {volume} {836}},\ \bibinfo {eid} {172} (\bibinfo {year}
  {2017})},\ \Eprint {http://arxiv.org/abs/1611.10292} {arXiv:1611.10292
  [astro-ph.HE]} \BibitemShut {NoStop}%
\end{thebibliography}%

\end{document}